\documentstyle[prd,aps]{revtex}
\begin{document}
\draft
\twocolumn[\hsize\textwidth\columnwidth\hsize\csname
@twocolumnfalse\endcsname
\preprint{SU-ITP-95-24, gr-qc/9511058}

\title{  ON REGULARIZATION SCHEME DEPENDENCE OF PREDICTIONS \\
IN   INFLATIONARY  COSMOLOGY}
\vskip 1cm
\author{ Andrei Linde
%\footnote{Electronic address: linde@physics.stanford.edu}
and Arthur Mezhlumian
%\footnote{Electronic address: arthur@physics.stanford.edu}
}
\address{Department of Physics, Stanford University, Stanford, CA
94305, USA}
\date{\today}
\maketitle
\begin{abstract}
We show that there exists a large class of regularization schemes for
probabilistic predictions in the theory of a self-reproducing inflationary
univere, all of which eliminate the apparent dependence
 on the time reparametrization. However, all these schemes lead to different
answers for relative probabilities of finding various types of
post-inflationary universes. Besides, all these schemes fail to be
reparametrization invariant beyond the range of the inflaton field close to end
of inflation boundary.  Therefore, we argue that at the current level of
understanding, the simple regularization schemes associated with cutoffs at
equal time hypersurfaces are as good as the recently proposed more complicated
procedures which try to fix the time-reparametrization dependence.
\end{abstract}
\pacs{PACS: 98.80.Cq, 98.80.Hw \hskip 5.6 cm SU-ITP-95-24,
{}~gr-qc/9511058}
\vskip2pc]

%%%%%%%%%%%%%%%%%%%%%%%%%%%%%%%%%%%%%%%%%%%%%%%%%%%%%%%%%%%%%%%
One of the most unusual, but quite general features of inflationary
models \cite{MyBook} is the  self-reproduction and
eternal expansion of the universe. This effect has especially dramatic
consequences in the context of chaotic inflation \cite{b19}, but in
fact it is very general and occurs in new inflation as well
\cite{Self-Reproduction}. For a wide range of initial conditions, the
universe enters a stage of stationary evolution which never ends, thus
making the very question of the original state almost irrelevant
\cite{LLM}.

The volume of the universe in such scenarios grows without end. In
particular, the volume of regions of the universe where inflation ends
and life becomes possible also grows without end. In many models of
inflation there exist several different minima of the effective
potential which may correspond to drastically different types of
post-inflationary physics: different values of constants of Nature,
different symmetry properties, and even different dimensionality
\cite{MyBook}. The number of different minima may be finite, but
sometimes we encounter the situations where there is a continuous
spectrum of all possible outcomes. A simplest example is given by the
inflationary Brans-Dicke cosmology, where the effective gravitational
constant after inflation may take all possible values from $0$ to
$\infty$ in different exponentially large parts of the universe
\cite{BD}. In such a situation the knowledge of the fundamental
Lagrangian by itself does not help us to explain the properties of the
part of the universe where we live, and one should invent a more
elaborate approach to explain observed properties of our universe or
to predict the result of the future observations.

There are various levels of ambition which may be taken when trying to
solve this problem. One may simply be satisfied by the fact that one
of the minima of the effective potential corresponds to the low-energy
physics of our type, invoking the weak anthropic principle
\cite{Anthropic} in order to explain why do we live in this particular
type of minimum rather than any other. In this approach we answer the
question whether it is {\em possible} to find ourselves in a local
part of the universe such as we observe around us, not attempting to
quantify how {\em probable} it is to find ourselves in there. This by
itself is a very significant progress. Indeed, eternal reproduction of
the universe in all its possible forms for the first time provides a
real justification for the use of the weak anthropic principle
\cite{MyBook,LLM,new_results}.\footnote{We should emphasize that the
  problem of choice between different vacua appears in a purely
  physical context and therefore it cannot be ignored even by those
  opposed to anthropic considerations for philosophical reasons.}

On the other hand, one may try to go even further and calculate
relative probabilities associated with various states of the universe
in order to prove that the state in which we happen to live is at
least not very improbable --- hoping to establish that we are in some
sense {\em typical}. In the absence of better understanding of the
origins of life, we may assume that the total number of observers
living in domains of the universe with some specific properties is
proportional to the total volume of such domains. Thus we presume that
life and observers do spontaneously originate if there is enough space
and if environment supports the life of that particular type.  Another
assumption is that we are typical observers, and therefore live in
those places where most other observers live.  (We will discuss the
validity of these assumptions in the end of the paper.)  Then by
finding the properties of the parts of the universe where typical
observers live one may try to explain some properties of our own part
of the universe.

This more advanced version of anthropic principle was extensively used
in inflationary cosmology for many years. For example, using this
approach it was possible to justify validity of the first version of
the Affleck-Dine mechanism of baryogenesis in the context of
inflationary cosmology \cite{AfflDine}. It was possible to show that
the standard constraints on the axion mass ($m_a {\
  \lower-1.2pt\vbox{\hbox{\rlap{$>$}\lower5pt\vbox{\hbox{$\sim$}}}}\ }
10^{-5}$ eV) disappear in inflationary cosmology if inflation ends
with the Hubble constant $H < 10^9$ GeV (e.g. in hybrid inflation)
\cite{Axions}. Several attempts have been made to obtain strong
anthropic constraints on the cosmological constant in the context of
inflationary cosmology \cite{CosmConst,VilMediocre,GBLinde,VM2}, to
find the most probable value of the gravitational constant and of the
parameter $\omega$ in the Brans-Dicke cosmology \cite{BD,BW}, etc.

The difference between the two versions of anthropic
principle is not clear cut. Although the weak anthropic principle does
not require calculation of the relative probability of observing the
local domain of the universe of our type as compared to domains of all
other types, if such probability turns out to be infinitesimally small
it would be difficult to justify the applicability of that principle.
Thus, the results of the second approach may invalidate the results of
the first one in some extreme cases. In general, one may consider the
second approach as an attempt of a more quantitative use of the weak
anthropic principle. Recently it was suggested to call this version of
anthropic principle ``the principle of mediocrity''
\cite{VilMediocre}, which emphasizes the assumption that we are
typical.  We will use a more conventional terminology for the reason to be
discussed in the end of this paper.

Calculation of probabilities ignoring self-reproduction of the
universe is a rather difficult but well defined task. Meanwhile, in an
eternally self-reproducing universe this problem becomes much more
complicated. Indeed, in this theory the universe looks like a growing
fractal consisting of indefinitely large number of regions of all
possible types.  Since the total volume of the self-reproducing
universe grows without end, the normalization of the probabilities
associated with volumes containing matter in specific states becomes
an ambiguous task.  Some sort of regularization of infinitely large
volumes should be made. If we try to compare volumes at different
equal time hypersurfaces, the dependence on time parametrization
proves to be very strong \cite{LLM,BD,GBLinde}. One is forced to make
a choice between different regularizations in order to make
quantitative predictions based on relative probabilities.

The importance of this ambiguity is emphasized by the fact that it is
not only the description of the global structure of the whole universe
(far beyond the visible scale) which depends considerably on the
choice of regularization. We found recently that many models of
inflation predict, under a specific choice of regularization (namely,
by making comparisons on the hypersurfaces of equal proper time), that
the local part of universe which we observe should be spherically
symmetric but may contain large radial inhomogeneity and we should be
living not far from the geometrical center of our local patch
\cite{LLMcenter,LLMcenter2}. Depending on the choice of an
inflationary model, this effect may be either insignificant and
practically unobservable \cite{BL} or very pronounced
\cite{LLMcenter}. Thus, the question of the dependence on time
parametrization, or more generally, dependence on the regularization
scheme in eternally inflating universe suddenly becomes related to
observations in a quite direct way.

In a recent article \cite{Vil_predict_1}  Vilenkin proposed a very
interesting new approach to regularization of diverging probabilities
in the context of eternal inflation. An important feature of his
approach is that it apparently eliminates the dependence of the
relative probabilities on the choice of the time parametrization. The
remaining weak dependence is argued to be of the same order as the
dependence on the operator ordering ambiguity in Wheeler--DeWitt
equation, i.\ e.\ within the accuracy of the quantum cosmology
approximation \cite{Vil_predict_2}.

However, it is not quite clear whether the absence of strong
dependence of the results on the choice of time parametrization is
enough to justify the use of this regularization procedure. We will
show in this article that, while retaining the general idea of
Vilenkin's regularization prescription, it is possible to find many
different implementations of this scheme. They do not coincide with
the implementation proposed in \cite{Vil_predict_1,Vil_predict_2},
and the answers which we get for relative probabilities are
considerably different from the answers obtained there. Nevertheless,
all those answers have the same time-reparametrization invariance
features as the ones proposed by Vilenkin.

We will show that the stationarity of inflation leads effectively to
implicit choice of time-parametrization in the regularization scheme
proposed in \cite{Vil_predict_1,Vil_predict_2}. Thus, the apparent
independence on time parametrization comes at a price of dependence on
regularization scheme.  We will show also that this scheme has
time-reparametrization invariance features for only part of the
universe, namely for very low-energy regions near the end of inflation
boundary and post-inflationary domains. The relative probabilities
regarding the domains which are still in the process of active
inflation do depend on time parametrization even in that approach.  We
will present some conclusions regarding the choices between various
non-invariant answers and whether one has to insist on finding
``better'' invariant answers for the type of questions which we are
interested in.

The evolution of the scalar field driving inflation (inflaton),
coarse-grained over the spatial region of size $H^{-1}$ (called
hereafter an $h$-region in short), in the slow rolling approximation
in arbitrary time parametrization is described by the stochastic
differential equation \cite{Star,GLM,new_results}:

\begin{eqnarray} \label{SDE_any_time}
   \frac{d \phi}{d\tau} & = & -\frac{V'(\phi )}{3H(\phi ) \, T(\phi)} +
           \frac{H^{3/2}(\phi )}{2 \pi \, T^{1/2}(\phi)} \, \xi
           (\tau) \nonumber \\
 & \equiv & v(\phi) + \sqrt{2 D(\phi)} \, \xi(\tau) ,
\end{eqnarray}
where $\xi (\tau)$ is a gaussian white noise,
$V(\phi)$ is the inflaton potential, and $H(\phi) = \sqrt{8 \pi \,
V(\phi) / 3 M_P^2} $ is the Hubble parameter during the slow roll.
Different time parametrizations are related to the proper time
by local path dependent transformation:

\begin{equation} \label{time_reparametrization}
          t \rightarrow \tau (t) = \int^{t} ds \, T(\phi_{\xi} (s))\ ,
\end{equation}
where $T(\phi)$ is a positive function, and its argument in
(\ref{time_reparametrization}) is a solution of (\ref{SDE_any_time})
under $T \equiv 1$ (proper time) with a particular realization of the
white noise. We also introduced compact notations (coinciding with
those in \cite{Vil_predict_2} for their set
$T(\phi)=H^{\alpha}(\phi)$ of time reparametrization functions) for
the inflaton field dependent drift velocity

\begin{equation}\label{drift}
v(\phi) = -\frac{V'(\phi )}{3H(\phi ) \, T(\phi)}\ ,
\end{equation}
and diffusion coefficient

\begin{equation}\label{std}
D(\phi) = \frac{H^3(\phi ) }{ 8 \pi^2 \, T(\phi) }\ .
\end{equation}

There is an ambiguity in interpreting equation (\ref{SDE_any_time})
(regardless of time variable chosen, even with $T \equiv 1$ which
corresponds to proper time). One can interpret it as a stochastic
differential equation in the sense of Ito or in the sense of
Stratonovich. Without delving deep into these details, we just note
that this ambiguity is related to operator ordering problem in
Wheeler--DeWitt approach to quantum cosmology
\cite{LLM,Vil_predict_2}. Since this source of ambiguity is not
essential in our present analysis, we will choose for definiteness the
multiplicative Stratonovich interpretation of (\ref{SDE_any_time}) in
this paper.

The local dynamics of inflaton field in the given $h$-region,
described by (\ref{SDE_any_time}), is accompanied by exponential
expansion of the physical volume which may be regarded as a branching
of $h$-regions into $e^3$ ``daughter'' domains within every interval
$\Delta t \sim H^{-1}$ of proper time. Each such domain continues
independent evolution according to (\ref{SDE_any_time}) with its own
realization of the white noise \cite{LLM}.  One can introduce various
probability measures describing this process.

The simplest measure is the probability distribution of finding the
value of inflaton $\phi$ at time $\tau$ in the given $h$-region
disregarding any new domains generated during the inflation:

\begin{equation}\label{define_P_c}
P_c(\phi, \tau) = \left\langle \delta \left( \phi_{\xi}(\tau) - \phi
\right) \right\rangle_{\xi} \ ,
\end{equation}
where $\phi_{\xi}(\tau)$ is the solution of (\ref{SDE_any_time}) with
a particular realization of the white noise $\xi(\tau)$, and the average is
taken over all possible realizations of $\xi(\tau)$. This distribution
satisfies the Fokker-Planck equation:

\begin{eqnarray}\label{FP}
\frac{\partial}{\partial \tau} P_c(\phi, \tau) & = &
        \frac{\partial}{\partial \phi} \left( D^{1/2}(\phi)
                \frac{\partial}{\partial \phi} \left(D^{1/2}(\phi)
                P_c(\phi,\tau) \right) \right) \nonumber \\
       &  -  & \frac{\partial}{\partial \phi} \left(v(\phi) P_c(\phi, \tau)
\right) \ .
\end{eqnarray}

However, as we argued above, the comoving probability measure
$P_c(\phi, \tau)$ is often not the best choice for calculating
probabilities regarding typical observers of our type. A different
measure, proportional to the physical volume generated during
inflation

\begin{eqnarray}\label{define_P_p}
P_p(\phi, \tau) & =  & \left\langle \delta \left( \phi_{\xi}(\tau) - \phi
\right) \,
\mbox{Volume}_{\xi}(\tau) \right\rangle_{\xi} \nonumber \\
& = & \left\langle \delta \left( \phi_{\xi}(\tau) - \phi \right) \, e^{ 3
\int_{0}^{\tau} \frac{H(\phi_{\xi}(\tau'))}{T(\phi_{\xi}(\tau'))}
d\tau' } \right\rangle_{\xi}\ ,
\end{eqnarray}
was introduced in \cite{GLM,Nambu,Mijic,LLM}. Its evolution is
described by an equation, which can be interpreted as a Fokker-Planck
equation for branching diffusion \cite{MezhMolch,LLM}:

\begin{eqnarray}\label{BranchFP}
\frac{\partial}{\partial \tau} P_p(\phi, \tau) & = &
         \frac{\partial}{\partial \phi} \left( D^{1/2}(\phi)
                \frac{\partial}{\partial \phi} \left(D^{1/2}(\phi)
                P_p(\phi,
\tau) \right) \right) \nonumber \\
       & - & \frac{\partial}{\partial \phi} \left(v(\phi) P_p(\phi, \tau)
\right) +  n(\phi) P_p(\phi, \tau)\ .
\end{eqnarray}
The intensity of branching is related to local rate of expansion
of the universe \cite{LLM}:

\begin{equation}\label{branch_intensity}
n(\phi) = \frac{3H(\phi)}{T(\phi)}\ .
\end{equation}

It was proven in \cite{LLM} that the evolution of the eternal
inflationary universe very soon approaches a stationary state, where
all probability measures are dominated at late times by their ground
state (fastest increasing or slowest decreasing) eigenfunctions:

\begin{equation}\label{stationary}
P(\phi, \tau) = \sum_{k=1}^{\infty} e^{\lambda_{k} \, \tau} \,
\psi_{k}(\phi)
\approx e^{\lambda_1 \, \tau} \, \psi_1(\phi)  .
\end{equation}
 Here $\psi_{k}(\phi)$ are the eigenfunctions of the diffusion
generating operators on the right hand side of (\ref{FP}) or
(\ref{BranchFP}), respectively, and $\lambda_{k}$ are the
corresponding eigenvalues, $\lambda_1 > \lambda_2 > \ldots > \lambda_k >
\ldots$. We have
estimated in \cite{LLM} that this asymptotic regime establishes itself
very quickly, within few thousands of Planck times.

Looking at equations (\ref{drift}) -- (\ref{branch_intensity}) it
becomes clear why most of the quantities calculated with the help of
either $P_c$ or $P_p$ will depend on the choice of time
parametrization (i.\ e.\ function $T(\phi)$ in our case). All parts of
these equations explicitly depend on $T(\phi)$, therefore by taking a
different functional form of $T(\phi)$ one will inevitably end up with
quite different solutions $P_c$ and $P_p$.

In previous papers \cite{LLM,LLMcenter} we argued that this
dependence is an inherent feature of our problem since we are dealing
with super-horizon scale quantities which need not be
gauge-independent. As we mentioned before, the stationary solution of
(\ref{BranchFP}) increases with time indefinitely, so we need to
perform a regularization to obtain relative probabilities. The
simplest cut-off procedure which we implemented by taking the
equal time hypersurfaces and comparing the volumes on such
hypersurfaces, although clearly not unique, has the advantage of being
directly related to the derivation of (\ref{BranchFP}) and the meaning
of $P_p(\phi, \tau) $ which is the portion of the physical volume
filled with inflaton at value $\phi$ on the time hypersurface $\tau$.

Moreover, it does not really make any difference if we count the
integral of the three-dimensional volume from $\tau=0$ up to the time
hypersurface $\tau$ or just count the volume in that hypersurface.
Since the volume (and $P_p$) increases with time in the stationary
regime as an exponent with positive constant coefficient $P_p(\phi,
\tau) \sim \psi_1^{(p)}(\phi) \exp(\lambda_1^{(p)} \, \tau)$, then the
integral over time is dominated by its upper limit and is proportional
(up to the pre-exponential constant factor $1/\lambda_1^{(p)}$) to the
integrand at the upper limit. Thus, the ratio of volumes or
probabilities calculated in these two ways coincide with exponential
accuracy.

The value of the constant $\lambda_1$ as well as the functional form
of the normalized stationary probability distribution $\psi_1(\phi)$
depend crucially on the choice of the time-parametrization (functional
form of $T(\phi)$). One of the few invariant features of the
probability distribution is whether it is exponentially increasing or
decreasing, or equivalently, whether there is a self-reproduction
manifest in this measure or not. In branching diffusion terminology,
the invariant feature is whether the branching process is
super-critical or sub-critical. This is determined by the sign of
$\lambda_1$, which does not change for most reasonable
reparametrizations (unless we choose ``geodesically incomplete''
parametrization \cite{LLM,Disser}).

The basic idea of Vilenkin's regularization scheme
\cite{Vil_predict_1,Vil_predict_2} is to use the comoving measure
$P_c$ for calculating the cut-off times, which are defined as times
when all but a small portion of the comoving volume is already
thermalized in the given minimum, and then to use those cut-off times
to calculate the regularized ratios of volume weighted probabilities
$P_p$. The reason why this works is that the comoving measure $P_c$
does not exhibit self-reproduction and its value decays with time
exponentially with a negative constant coefficient $P_c(\phi, \tau)
\sim \exp(\lambda_1^{(c)} \tau)$ due to the continuous outflow of
probability into post-inflationary stage. In this measure any
$h$-region slowly but steadfastly rolls down towards the values of
inflaton field when the inflationary approximation breaks and the hot
big bang begins.  One can prove that $\lambda_1^{(c)} < 0$ if there is
any kind of boundary where inflationary regime ceases to exist.
Therefore, the total probability $P_c$ of finding a domain still in
inflationary stage converges at $t \rightarrow \infty$ and one can
make a well defined statement on what portion of comoving volume is
already in post-inflationary state.

However, from what is said above it is clear that there is nothing
special about the comoving volume measure $P_c$ (besides the fact that
it is well known and studied) --- any measure which converges at $\tau
\rightarrow \infty$ will be suited just as well for purposes of
regularization. For example, one can consider a probability weighted
with some measure not coinciding with volume. A particular case would
be a probability density associated with branching diffusion process
where branching intensity (in proper time) is just some arbitrary
function $n(\phi) \neq 3 H(\phi)$ unrelated to the local rate of
expansion of the universe. Since any branching intensity scales with
time reparametrizations as $1/T(\phi)$, the cancellation of the time
parametrization dependence in the regularized volume ratio will still
hold, as one can easily see from the final results
below\footnote{There is a particular choice of $n(\phi)$ which allows
  a simple interpretation. Let us take $n(\phi)=C/T(\phi)$, which
  becomes simply $n(\phi)=C$ for proper time. Then eq.
  (\ref{BranchFP}) describes the distribution of such observers who
  split producing a constant average number $C$ of new observers per
  unit proper time. For $C < 0$, (\ref{BranchFP}) describes observers
  who have a finite proper lifetime and, in average, die after $\Delta
  t \sim C^{-1}$.  }.

To be specific, we suggest the following, slightly more general
measure to be used instead of $P_c$ in the cut-off procedure while
retaining the general idea of \cite{Vil_predict_1,Vil_predict_2}:

\begin{eqnarray}\label{define_P_q}
P_q(\phi, \tau) & = & \left\langle \delta \left( \phi_{\xi}(\tau) - \phi
\right) \,
\left(\mbox{Volume}_{\xi}(\tau) \right)^q \right\rangle_{\xi} \nonumber \\
& = & \left\langle \delta \left( \phi_{\xi}(\tau) - \phi \right) \, e^{ 3 \, q
\,
\int_{0}^{\tau} \frac{H(\phi_{\xi}(\tau'))}{T(\phi_{\xi}(\tau'))}
d\tau' } \right\rangle_{\xi}\ .
\end{eqnarray}

This measure can be considered as appearing from multi-fractal
analysis of the global structure of inflationary universe. Indeed, it
has been established that the geometry of the eternal inflationary
universe is fractal \cite{GLM,ArVil,LLM}. However, different measures
lead to different fractal dimensions. A natural way to interpret this
structure is to consider the physical volume associated with every
point as a multi-fractal measure defined on the underlying comoving
base geometry. Then, the characterization (\ref{define_P_q}) is very
similar to the way the so called Renyi dimensions are introduced
\cite{Fractal}. Physically speaking, this measure corresponds to
taking into account only a fraction of newly generated volume at every
time step. Taking $q=0$ we return to the case of $P_c$, while $q=1$
corresponds to $P_p$ (we will use superscript notations $c
\leftrightarrow (q=0)$ and $p \leftrightarrow (q=1)$ for compactness
and to retain compatibility with our notations in
\cite{LLM})\footnote{The correspondence with notations in
  \cite{Vil_predict_1,Vil_predict_2} is established using the
  following rules: their quantities without tilde correspond to our
  quantities with superscript $(c)$, the quantities with tilde
  correspond to our ones with superscript $(p)$, their $\gamma_k$
  corresponds to our $\lambda_k$, while $\tilde{\gamma}
  \leftrightarrow \lambda_1^{(p)}$ and $\gamma \leftrightarrow
  -\lambda_1^{(c)}$.}. In general, one may consider $q < 0$ as well.

This {\it multi-fractal} measure satisfies a branching diffusion   equation,
where the branching intensity is proportional to the old value
(\ref{branch_intensity}) with a coefficient $q$:

\begin{eqnarray}\label{FractalFP}
\frac{\partial}{\partial \tau} P_q(\phi, \tau) & = &
         \frac{\partial}{\partial \phi} \left( D^{1/2}(\phi)
                \frac{\partial}{\partial \phi} \left(D^{1/2}(\phi)
                P_q(\phi, \tau) \right) \right) \nonumber \\
       & - & \frac{\partial}{\partial \phi} \left(v(\phi) P_q(\phi, \tau)
\right) +  q n(\phi) P_q(\phi, \tau)\ .
\end{eqnarray}

It is clear that for small enough values of $q$ the probability distribution
$P_q$  (\ref{FractalFP})  will be
decaying with time just like $P_c$. Therefore, for small enough values
of $q$ this measure can be used as an alternative regularization
measure. Following the prescriptions in
\cite{Vil_predict_1,Vil_predict_2}
we rewrite (\ref{FractalFP}) in a probability flux form:

\begin{equation}\label{FluxFP}
\frac{\partial}{\partial \tau} P_q(\phi, \tau) =
        - \frac{\partial}{\partial \phi} J_q(\phi, \tau)
        +  q n(\phi) P_q(\phi, \tau)\ ,
\end{equation}

\begin{eqnarray}\label{define_flux}
J_q(\phi, \tau) & = & -  D^{1/2}(\phi)
                \frac{\partial}{\partial \phi} \left(D^{1/2}(\phi)
                P_q(\phi, \tau) \right) \nonumber \\
 &+ & v(\phi) P_q(\phi, \tau) \ .
\end{eqnarray}

Let there be two types of post-inflationary minima, denoted
$\phi_{ei}^{(1)}$ and $ \phi_{ei}^{(2)}$, respectively, where the
subscript stands for ``end of inflation''. The
total volume thermalized in these minima by time $\tau$ is given by
the integral of the volume flux through the corresponding boundary:

\begin{equation} \label{volume_flux}
{\cal V}^{(1,2)}=\left| \int\limits_{0}^{\tau} J_{p}\left(
\phi_{ei}^{(1,2)}, \tau' \right) d\tau' \right| \ ,
\end{equation}
where the symbol (1,2) is used for the first or the second minimum
correspondingly.

The  volumes ${\cal V}^{(1)}$ and ${\cal V}^{(2)}$ diverge at large $\tau$.
Vilenkin's regularization prescription
is to introduce cut-off times $\tau_{\epsilon}^{(1,2)}$ such that only a small
portion of {\it converging} measure is still in the inflationary stage. We will
use a general multi-fractal measure $P_q$ with small enough $q$ as such
converging measure:

\begin{equation} \label{cut_off}
\left| \int\limits_{\tau_{\epsilon}^{(1,2)}}^\infty J_{q}\left(
\phi_{ei}^{(1,2)}, \tau\right) d\tau \right|   =
\epsilon \left| \int\limits_{0}^\infty J_{q} \left(
\phi_{ei}^{(1,2)}, \tau\right) d\tau \right|    \equiv   \epsilon p^{(1,2)}  .
\end{equation}
Here $p^{(1)}$ and $p^{(2)}$ are the total multi-fractal measures that
will eventually thermalize in the first and second minima
respectively. Then, the regularized expressions for total physical
volume which will end up in the corresponding minima are given by:

\begin{equation} \label{define_volume_regularized}
{\cal V}^{(1,2)}=\left| \int\limits_{0}^{\tau_{\epsilon}^{(1,2)}}
J_{p}\left(\phi_{ei}^{(1,2)}, \tau\right) d\tau\right|\ .
\end{equation}

In order to estimate  these integrals we have to find the flux
(\ref{define_flux}) near the end of inflation boundary.  Fortunately, it
is a well known fact that below certain level of energy density one
can neglect the diffusion (second derivative) part in
(\ref{FractalFP}) because it is much smaller than the drift (first
derivative) part \cite{MyBook,GLM}. Therefore, within the
applicability of this approximation in the stationary regime the flux
is:

\begin{equation}\label{flux_near_end}
J_q(\phi, \tau) \approx v(\phi) P_q(\phi, \tau) \approx
v(\phi) \, \psi_1^{(q)}(\phi) \, e^{\lambda_1^{(q)} \, \tau} .
\end{equation}

For small enough $\epsilon$ the characteristic cut-off times become
very large and the use of stationary asymptotic (\ref{stationary}) is
justified. Indeed, the volume flux integral
(\ref{define_volume_regularized}) is dominated by its upper limit,
while the left hand side of (\ref{cut_off}) is dominated by its lower
limit because of the signs of exponential coefficients
$\lambda_1^{(p)} > 0$ and $\lambda_1^{(q)} < 0$ for sufficiently small
$q$. As we mentioned before, for any $q$ (including $q=1$)
$\lambda_1^{(q)}$ is the highest eigenvalue associated with the right
hand side of (\ref{FractalFP}):

\begin{eqnarray}\label{StationaryFP}
\lambda_1^{(q)} \psi_1^{(q)}(\phi) & = &
         \frac{d}{d \phi} \left( D^{1/2}(\phi)
                \frac{d}{d \phi} \left(D^{1/2}(\phi)
                \psi_1^{(q)}(\phi) \right) \right) \nonumber \\
       & - & \frac{d}{d \phi} \left(v(\phi) \psi_1^{(q)}(\phi)
\right) +  q n(\phi) \psi_1^{(q)}(\phi)\ .
\end{eqnarray}
Substituting (\ref{flux_near_end}) into (\ref{cut_off}) and
(\ref{define_volume_regularized}), we get the cut-off times:

\begin{equation} \label{cut_off_times}
\tau_{\epsilon}^{(1,2)} =  \frac{1}{\left| \lambda_1^{(q)} \right|}
    \ln\left(\frac{\left| v(\phi_{ei}^{(1,2)}) \right| \,
       \psi_1^{(q)}(\phi_{ei}^{(1,2)})}{\epsilon \, \left|
        \lambda_1^{(q)} \right| \,
    p^{(1,2)}} \right) ,
\end{equation}
and the regularized volumes:

\begin{equation} \label{volume_regularized}
{\cal V}^{(1,2)} = \frac{1}{\lambda_1^{(p)}} \,
    \left| v(\phi_{ei}^{(1,2)}) \right| \,
    \psi_1^{(p)}(\phi_{ei}^{(1,2)}) \,
e^{\lambda_1^{(p)} \,  \tau_{\epsilon}^{(1,2)}}  .
\end{equation}
{}From (\ref{cut_off_times}) and (\ref{volume_regularized}) we get the
regularized ratio of volumes thermalized in the  two minima:

\begin{eqnarray} \label{ratio}
r  \equiv   \frac{{\cal V}^{(2)}}{{\cal V}^{(1)}}
 & = & \frac{\left| v(\phi_{ei}^{(2)}) \right| \, \psi_1^{(p)}
(\phi_{ei}^{(2)})}{
    \left| v(\phi_{ei}^{(1)}) \right| \,  \psi_1^{(p)} (\phi_{ei}^{(1)})}
\nonumber \\
   & \times & \left( \frac{\left| v(\phi_{ei}^{(2)}) \right| \,  \psi_1^{(q)}
      (\phi_{ei}^{(2)})}{
    \left| v(\phi_{ei}^{(1)}) \right| \,  \psi_1^{(q)}
    (\phi_{ei}^{(1)})} \;
     \frac{p^{(1)}}{p^{(2)}}\right)^{
       \frac{\lambda_1^{(p)} }{\left| \lambda_1^{(q)} \right|}} .
\end{eqnarray}
This result is a generalization of equation (41) from article
\cite{Vil_predict_2} for the multi-fractal measure $P_q$ used instead
of $P_c$ for defining the cut-off procedure.  Within the applicability
of the no-diffusion approximation used to derive (\ref{ratio}) one can
easily solve (\ref{StationaryFP}):

\begin{equation}\label{eigen_function}
\psi_1^{(q)}(\phi) = \frac{C_1^{(q)}}{v(\phi)} \,
        \exp\left( \int\limits_{\phi_0}^{\phi}
                \left[ q \frac{n(\phi')}{v(\phi')}-
\frac{\lambda_1^{(q)}}{v(\phi')}
\right]
d\phi' \right)\ .
\end{equation}

We can see immediately that the dependence on the functional form
$T(\phi)$ cancels out in the first term in the exponent, therefore it
is time reparametrization invariant. The second term depends on
$T(\phi)$, and therefore it is crucial that this term cancels in the
formula for ratio of volumes (\ref{ratio}). It indeed works out very
nicely  (remember that $\lambda_1^{(q)} < 0$ for small $q$), and
the final result after substituting (\ref{eigen_function}) into
(\ref{ratio}) becomes:

\begin{equation}\label{ratio_final}
r = C(q) \, \left( \frac{Z^{(2)}}{Z^{(1)}} \right)^3\ ,
\end{equation}
where the mildly varying coefficient $C(q) \sim 1$ is not
important\footnote{There is a subtle point here which is worth
  mentioning. The values of total multi-fractal measures $p^{(1)}$ and
$p^{(2)}$
  which will eventually thermalize in the two minima are {\it not}
  dependent on time parametrization. This is because regardless of
  which time variable we choose, the physics of the result remains the
  same for all converging measures: every given $h$-region will sooner
  or later be either in the first or in the second minimum. Therefore when
  calculating the total thermalized converging measure one will always
  get the same number.}, and the exponentially large volume factors
are:

\begin{eqnarray}\label{volume_factors}
Z^{(1,2)} & = & \exp\left\{\left(1 + q \frac{\lambda_1^{(p)}}{\left|
\lambda_1^{(q)} \right|} \right) \,  \int\limits_{\phi_0}^{\phi_{ei}^{(1,2)}}
\frac{n(\phi)}{3 \, v(\phi)} d\phi \right\} \nonumber \\
& = & \exp\left\{ -\left(1 + q \frac{\lambda_1^{(p)}}{\left|
\lambda_1^{(q)}
\right|} \right) \,  \int\limits_{\phi_0}^{\phi_{ei}^{(1,2)}}  \frac{
 4 \pi \, H(\phi)}{H'(\phi)} d\phi \right\}\ .
\end{eqnarray}

In the case $q = 0$ this result   coincides with   the one obtained by Winitzki
and Vilenkin
\cite{Vil_predict_2}, see their  eq.\ (46). However,   for small but not
vanishing $q$, our result is exponentially different. As it is
obvious from (\ref{volume_factors}), our result has the same time
reparametrization invariance features as its particular case in $q=0$,
since it does not explicitly depend on the function $T(\phi)$, while
the ratio of eigenvalues is stable with respect to time
reparametrization, as it was shown in \cite{Vil_predict_2}. Therefore,
we have proved that there exists a large family of regularization
schemes, all of which satisfy the reparametrization invariance
condition, but all lead to considerably different from each other
answers.

Our results suggest that even nice features like time
reparametrization invariance are not enough to uniquely determine the
{\it correct} way of regularization of the quantities in quantum
cosmology. It is not clear why would one of our multi-fractal
regularization schemes be any better or worse than a particular scheme
associated with $P_c$.  One indeed has to have a much deeper than
purely technical reasons in order to pick a particular regularization
scheme in favor of many others.

Another important note regarding the reparametrization invariance of
the regularized volume ratio is that its invariance is only partial.
Indeed, the expression (\ref{ratio_final}), and even the whole
expression (\ref{ratio}) are valid only sufficiently close to end of
inflation boundaries, where the no-diffusion approximation holds. Beyond
the region of its applicability, the explicit dependence on the
functional form of $T(\phi)$, and therefore dependence on time
reparametrizations will be unavoidable, since the more complicated
form of the complete solution of branching diffusion equation
\cite{LLM,Disser} does not allow for magic cancellations like the one
in (\ref{ratio_final}).

This is even more important for chaotic inflation models most of which
have a very narrow range of values of the inflaton field where the
no-diffusion approximation is valid.  For example, in the simplest
such model based on the inflaton potential $\lambda \phi^4/4$ the
total range of the possible values of the inflaton field is $0 {\
  \lower-1.2pt\vbox{\hbox{\rlap{$<$}\lower5pt\vbox{\hbox{$\sim$}}}}\ }
\phi {\
  \lower-1.2pt\vbox{\hbox{\rlap{$<$}\lower5pt\vbox{\hbox{$\sim$}}}}\ }
\lambda^{-1/4} M_{\rm P} \sim 2\times10^{3}\, M_{\rm P}$ (here
$\lambda \sim 10^{-13}$ is the value of the coupling constant obtained
from constraints on CMB anisotropy).  At the same time the range of
applicability of the no-diffusion approximation is only $0 < \phi <
\lambda^{-1/8} M_{\rm P} \sim 40\, M_{\rm P}$, i.~e.\ narrower by
almost two orders of magnitude \cite{LLM,LLMcenter}.

Thus, in the simplest versions of chaotic inflation scenario the
procedure suggested in \cite{Vil_predict_1,Vil_predict_2} and its
generalizations proposed above fail to produce reparametrization
invariant results in the main part of the allowed values of the field
$\phi$. In terms of energy density $\rho$, the results become
reparamerization invariant only for $\rho$ about six orders of
magnitude smaller than the Planck density!  Of course, one could argue
that predictions of quantum cosmology need not be reparametrization
invariant during inflation since life cannot appear there anyway.
However, this does not sound like a satisfactory answer if one want to
maintain the consistency of this approach.

Despite the problems mentioned above, one should note that the
regularization procedure suggested by Vilenkin has some advantages,
and its results have a very simple physical interpretation: the ratio
of volumes (\ref{ratio_final}) in the case $q = 0$ coincides with its
value in the classical theory ignoring the process of
self-reproduction. Indeed, the factor
\begin{equation}\label{classical}
 Z^{(1)}  =  \exp\left( - 4 \pi
   \int\limits_{\phi_0}^{\phi_{ei}^{(1)}}
   \frac{H(\phi)}{H'(\phi)} d\phi \right)\
\end{equation}
coincides exactly with the total expansion factor of the universe
during the classical rolling of the field $\phi$ from $\phi_0$ to
$\phi_{ei}^{(1)}$.

However, one may consider this result as being too simple: the central
feature of our scenario, the effect of self-reproduction of the
universe, has completely dropped out from the answer (for $q = 0$).
This is somewhat disturbing, but technically it is quite
understandable since the regularization scheme of
\cite{Vil_predict_1,Vil_predict_2} is based on the investigation of
the distribution $P_c$, which is not very suitable for study
of the self-reproduction. This distribution correctly describes the
typical fate of each individual observer which was present in the
universe from the very beginning. However, there were no real observers
during inflation, and the whole observable part of our universe was
created from the place which was much smaller than the Planck length
until the last 70 e-foldings of inflation. Thus, inhabitants of our
part of the universe could not care less about the fate of imaginary
observers living at the beginning of inflation.  By focusing our
attention on the portion of such ``observers''   entering the
post-inflationary epoch we are ignoring the main portion of the volume
of the universe which is produced in those rare domains which are
jumping back towards large $V(\phi)$ in the process of
self-reproduction of the universe.

Note also that all regularization procedures based on the use of the
convergent measure $P_q$ have one feature in common. Since for any
time parametrization the eigenvalues $\lambda_1^{(q)}$ are constants,
it is evident from (\ref{cut_off_times}) and
(\ref{volume_regularized}) that the regularization procedure
considered in this paper implicitly favors a specific time
parametrization, namely the parametrization in which time variable is
the logarithm of the total volume (this variable was first introduced
in the context of stochastic inflation in \cite{Star}). Indeed, the
stationarity of inflation leads both the converging measure and
diverging volume to be proportional to exponents of time variable with
different coefficients. Therefore, cut-off times
$\tau_{\epsilon}^{(1,2)}$ are proportional to logarithms of volume,
making this choice of time parametrization implicitly preferred. In
fact, it was argued in \cite{Vil_predict_1} that this time
parametrization is indeed preferred, and the scale factor was the only
variable to be used as clock during inflation. This is very similar to
the argument used by Hawking when he was trying to prove that the
arrow of time should turn back at the moment when the scale factor of
a closed universe begins decreasing \cite{Arrow}. The main problem
with this argument is that the scale factor of the universe is the
best clock only before the universe is observed. Then its wave
function collapses into the wave function depending on the clock used
by an observer, and after that the decrease of the scale factor does
not lead to the reversal of the new time arrow \cite{PageHawk}. The
natural clock used by observers of our type are based on oscillators
rather than on the measurement of distances between galaxies.

Thus we need some additional reasons to pick up a particular
regularization and use it to define a proper measure in quantum
cosmology.  Such reasons might not be completely determined by quantum
field theoretical consistency of the description of eternally
inflating universe, e.g. by time reparametrization invariance. One may
need to add additional considerations regarding the type of questions
which we are asking. In particular, one might want to take into account
the preferences for regularization introduced by the fact that we, the
observers, are like what we are. Although the scheme suggested in
\cite{Vil_predict_1,Vil_predict_2} is indeed very appealing in its
logic, that logic is not necessarily the correct one from the point of
view of the observers like ourselves who live and evolve very much
along the proper time clocks rather than anything else, and that by
itself may be the reason to pick the proper time regularization
\cite{LLM}. Since we ourselves are not gauge invariant objects with
respect to time reparametrization, it may be not so bad to have some
of the results concerning the probability of creation of domains of
our type to be reparametrization dependent too.

The probability distribution $P_p(\phi,t)$ used in our previous works
has certain advantages as it comes most closely to the description of
the total volume of the universe at a given time (including our
present time).  Therefore we are going to continue its investigation
in our subsequent publication \cite{LLMcenter2}. However, it is quite
possible that there may be better suggestions.  One of the ideas for a
physically motivated cut-off can be explained as follows. We speak
about the new regions of the universe produced during inflation as of
the regions of  classical space-time. Meanwhile they are produced by
quantum fluctuations. Classical description of the new regions is
indeed quite legitimate since they rapidly grow
exponentially large.  However, they are not {\it infinitely }
large. There is some time after which the wave function describing
different parts of the universe will cease describing classical
trajectories and will begin interfere quantum mechanically.  We can
easily estimate the size of a locally homogeneous part of our universe
(${\delta\rho\over \rho} \sim 1$) and the typical time before
gravitational collapse of a typical locally Friedmannian part of the
universe.  This might be the time when the physical cut-off should be
introduced. It is not quite clear to us that this is the right way to
think about this problem, but it would be very interesting to pursue
this idea.

At the present moment we cannot say which of the ways to introduce
measure in inflationary cosmology is really preferable. It is possible
that eventually we will resolve this problem. Still even that will not
guarantee that we are on the right track.  Our investigation was based
on two hidden assumptions. The first assumption is that we are typical
observers. The second assumption is that the number of typical
observers is directly proportional to the volume of the universe.  If
this is correct, then we should live in the place where most observers
live, which should correspond to the maximum of the probability
distribution.

However, is it not absolutely clear whether the probability for an
observer to be born in a particular part of the universe is directly
proportional to its volume, or one should take into account something
else. One cannot get any crop even from a very large field without
having seeds first. The idea that life appears automatically once
there is enough space to be populated may be too primitive.  It is
based on the assumption that one can describe emergence of life solely
in terms of physics, and that everything which is necessary for
emergence of life was created {\it after} inflation.  It is certainly
the most economical approach, and one should try to go as far as
possible without invoking additional hypotheses. One should keep in
mind, however, that this approach may happen to be incomplete,
especially if consciousness has its own degrees of freedom
\cite{MyBook,Page}.  The nature of the problem can be formulated in
purely physical terms as well.  Suppose, for example, that emergence of
life is catalyzed by the presence of some particles which, like
primordial monopoles, cannot be produced at the late stages of
inflation. Then the growth of volume of the universe at the late
stages of inflation does not increase the total number of such
particles and, consequently, the number of observers to be born. In
such a situation none of the regularization procedures discussed in
the present paper will give a correct information about the parts of
the universe where typical observers live.

Another related question is whether we are actually typical? Does it
make any sense for each of us to calculate {\it a posteriori} the
probability to be born Russian, Italian or Chinese? Should we insist
on our own mediocrity, or, {\it vice versa}, should we try to explain
why are we so special? After all, for a long time we thought that we
had an aristocratic privilege to be the most intelligent species in
the universe. This, of course, may be wrong.  Still, before using
probabilities to calculate the likelihood of our existence in a
particular part of the universe, it may be a good idea to learn more
about ourselves.  It might happen that until we understand what is our
life and what is the nature of consciousness our understanding of
quantum cosmology will remain fundamentally incomplete.

In the meantime one may take a pragmatic point of view, postpone
answering all these questions and consider our investigation as a kind
of ``theoretical experiment.''  We may try to use probabilistic
considerations in a trial-and-error approach. If we get unreasonable
results, this may serve as an indication that we are using quantum
cosmology incorrectly. However, if some particular proposal for the
probability measure in quantum cosmology will allow us to solve
certain problems which could not be solved in any other way, then we
will have a reason to believe that this choice of the probability measure is
preferable and that perhaps we are moving in the right direction.

We are grateful to A. Vilenkin for many enlightening discussions. This
work was supported by NSF grant PHY-8612280.

\end{document}